# Random Linear Network Coding for Non-Orthogonal Multiple Access in Multicast Optical Wireless Systems

Ahmed Ali Hassan[1], Ahmed Adnan Qidan[2], Taisir Elgorashi[1], Jaafar Elmirghani[1]

[1]Departement of Engineering, King's College London, WC2R 2LS London, UK
[2]School of Electronic and Electrical Engineering, University of Leeds, LS2 9JT Leeds, UK
e-mail: {Ahmed.4.Hassan@kcl.ac.uk, A.A.Qidan@leeds.ac.uk, Taisir.Elgorashi@kcl.ac.uk, Jaafar.Elmirghani@kcl.ac.uk}

**ABSTRACT**
Optical Wireless Communication networks (OWC) has emerged as a promising technology that enables high-speed and reliable communication bandwidth for a variety of applications. In this work, we investigated applying Random Linear Network Coding (RLNC) over NOMA-based OWC networks to improve the performance of the proposed high density indoor optical wireless network where users are divided into multicast groups, and each group contains users that slightly differ in their channel gains. Moreover, a fixed power allocation strategy is considered to manage interference among these groups and to avoid complexity. The performance of the proposed RLNC-NOMA scheme is evaluated in terms of average bit error rate and ergodic sum rate versus the power allocation factor. The results show that the proposed scheme is more suitable for the considered network compared to the traditional NOMA and orthogonal transmission schemes.

## 1. INTRODUCTION

The increasing demand of high-speed and reliable data communication in wireless networks that efficiently distribute content to multiple users has gained significant attention in variety of applications such as video streaming, edge computing, and online gaming. In context of Optical Wireless Communication (OWC), several challenges are encountered in achieving higher data rates using conventional optical transmitters and receivers such as limited modulation bandwidth, inter-channel-interference, and multipath dispersion[1]–[4]. To increase the system capacity and enable multiple users communication, various orthogonal techniques that have been applied in RF networks are also investigated in OWC networks i.e., Time Division Multiple Access (TDMA), Frequency Division Multiple Access (FDMA), and Code Division Multiple Access (CDMA). these schemes provide inadequate performance in high density optical wireless networks as the resources are exclusively allocated to limited number of time slots, frequency bands or codes[5]–[8].

Non-orthogonal Multiple Access (NOMA) is a key promising technology for improving the spectral efficiency and system capacity of optical wireless networks. it allows multiple users to share the same time slot or frequency band by superimposing different power levels or codes assigned to users signals to enable simultaneous transmission. In the reception, the desired user signal can be obtained by applying Successive Interference Cancellation (SIC) on the other received users signals[9]–[12]. In power domain NOMA, users are ascendingly ordered by their channel gains or proximity from an optical transmitter. The higher power levels are allocated to the far or low channel gain users, and the low power levels are allocated to the near or high channel gain users. However, NOMA might suffer performance degradation which induces packets loss or service outage; This can be due to multiple factors such as outdated or partially outdated Channel State Information (CSI), imperfect SIC, and Inter Cell Interference (ICI) [13]–[17].

On the other hand, Random Linear Network Coding (RLNC) is a communication technique that enables efficient data transmission over unreliable communication channels by using algebra to combine the data packets together and transmits the generated combinations over the network[18]–[22]. Unlike the error-correcting coding in traditional communication systems which add redundancy to the data and resulting in inefficient spectrum usage, RLNC allows mixing the original packets together with randomly chosen coefficients from a finite Galois Field (GF)[23]. the coded packets are transmitted to users where gaussian elimination process is performed to recover the original packets. Several studies proposed RLNC in NOMA-based RF broadcast and multicast networks to overcome the packets loss and reduce the number of packet re-transmissions without the need of packet loss feedback or sequence synchronisation[24], [25]. In OWC systems, several network coded schemes were studied in relay-based optical wireless networks[26], [27] and optical physical links in core networks[28], [29]. In this work, we investigated applying RLNC with NOMA in indoor multicast OWC networks and analysed the performance of the considered network model in terms of average bit error rate (BER) and ergodic sum rate taking into consideration the imperfect SIC factor in our analysis.

The rest of the paper is organized as follows: The system model is presented in Section 2, In Section 3, the simulation results are discussed, Finally, the conclusions are given in Section 4.

## 2. SYSTEM MODEL

We consider a simulation model of OWC multicast network in an indoor environment as shown in Figure 1. The network model consists of a single optical transmitter with Lambertian radiation patten i.e., Light Emitting Diode (LED) that mounted in the middle of the room ceiling to serve multicast users that are uniformly distributed on the communication plane. Each user receiver is equipped with 1 cm$^2$ active area and 35º field of view (FOV) photo Diode (PD) that points to the ceiling. The users are divided into two multicast groups based on their channel gains, the first group includes the weak users, and the second group includes the strong users. It is assumed the LED transmitter is connected to a central control unit that controls the network resources and it has a complete knowledge of users CSI and users locations distribution at any given time. The line-of-sight (LoS) component is considered as a direct optical link between the LED source and each user and the reflections of the room walls or any diffuse objects are neglected for the sake of simplicity.

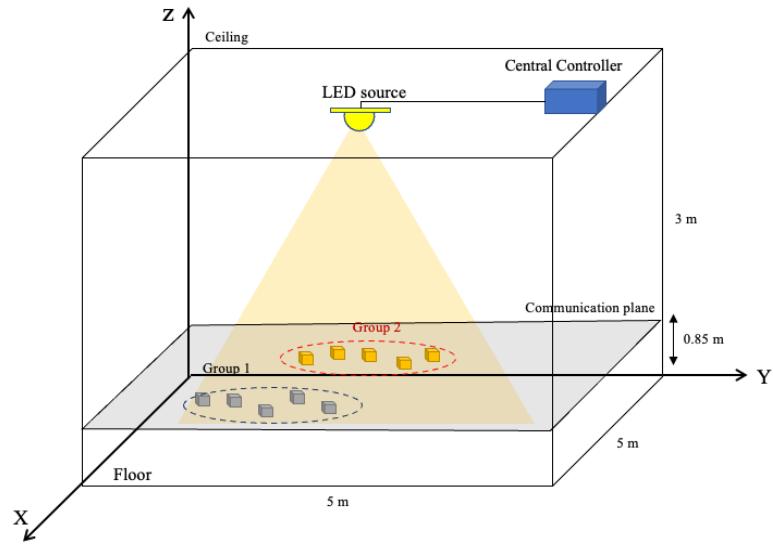

*Figure 1.OWC system model of two multicast groups*

NOMA principle is applied to align the interference between the two multicast groups, and two fixed distinctive power levels are assigned to the transmitted signals. The higher power level is allocated to weak users group (group 1) while the lower power level is allocated strong users group (group 2). The both selected power levels are constrained by the minimum and maximum LED driver current value to satisfy the illumination and communication requirements in typical indoor OWC systems[30]. A frame of source packets are combined with randomly chosen coding coefficients from Galois Field GF($2^8$).The generated set of coded packets are modulated using On-Off Keying (OOK) and assigned to the allocated power levels. The RLNC-NOMA scheme operations are illustrated in Figure 2., and the remaining system parameters are detailed in Table1.

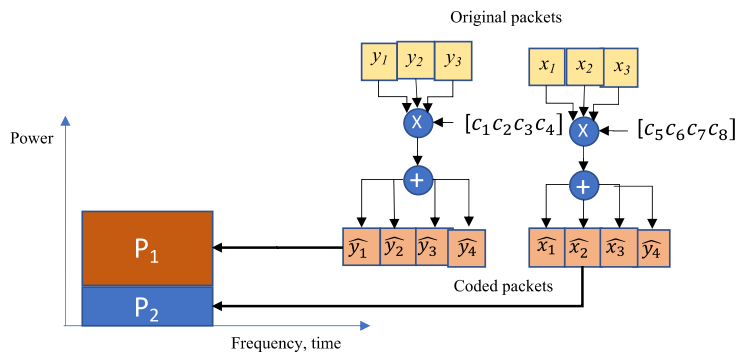

*Figure 2. RLNC-NOMA scheme operations*

*Table 1. system parameters.*

| Parameter | Value | Unit |
| --- | --- | --- |
| Room size | 5 x 5 x 3 | m$^3$ |
| Cell size | 3.6 | m |
| LED output power | 1 | W |
| Half-power angle | 60 | Deg. |

| LED Transmitter position | (2.5,2.5,3) | (x,y,z) |
|---|---|---|
| Users Height | 0.85 | m |
| Responsivity of PD | 0.4 | A/W |
| Number of users in each multicast group | 5 | – |
| Modulation bandwidth | 20 | MHz |
| Noise power spectral density | $10^{-21}$ | W/Hz |
| Number of packets in the RLNC frame | 10 | – |
| Minimum NOMA user throughput | 0.5 | bits/s.Hz$^{-1}$ |

## 3. SIMULATION RESULTS AND DISCUSSIONS

A Monte-Carlo simulation was developed using MATLAB to evaluate the performance of the proposed RLNC-NOMA scheme and compare it to the conventional NOMA and OMA schemes. A set of coded packets are randomly generated and transmitted to the users of the two multicast groups using the mentioned system parameters in Table1. A power allocation coefficient $\alpha$ is used to determine the allocated power ratio of both multicast groups. In Figure 3, the ergodic sum rates of RLNC in both groups are analysed, and the results are clearly showing the sum rates of NOMA outperformed the OMA scheme rates. Note that, the assigned power allocation α is associated to the multicast group 2 and the power ratio of the multicast group 1 is (1 - α).

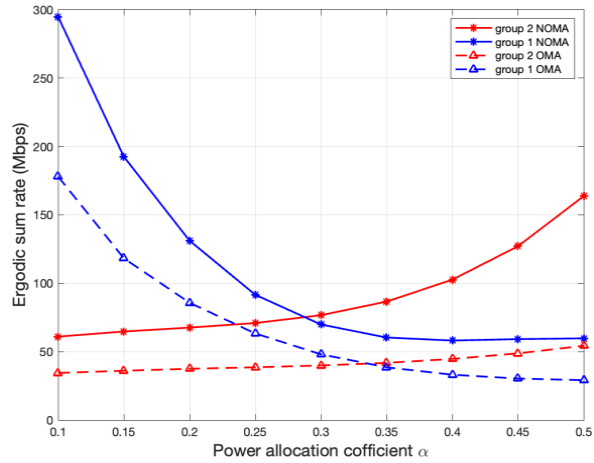

*Figure 3. NOMA and OMA ergodic sum rates versus power allocation coefficient α.*

The average BER versus the allocated power coefficient $\alpha$ results are depicted in Figure 4. It shows that RLNC-NOMA improves the average BER of the conventional NOMA in the case of imperfect SIC by 41% when $\alpha$ value is between 0.15 and 0.3 and the average signal-to-noise ratio (SINR) among the users in the same group is less than 12 dB. Note that, group 2 users perform SIC to decode the group 1 information and perform gaussian elimination to recover the original source packets. Moreover, the users of group 1 only decode their intended coded packets as the received signals of the group 2 are considered as a noise. Despite the significant improvement of RLNC-NOMA, the figure further shows the average BER increases when α factor increases and approaching the equal power allocation between the two multicast groups which is expected due to high interference between the signals of the two multicast groups.

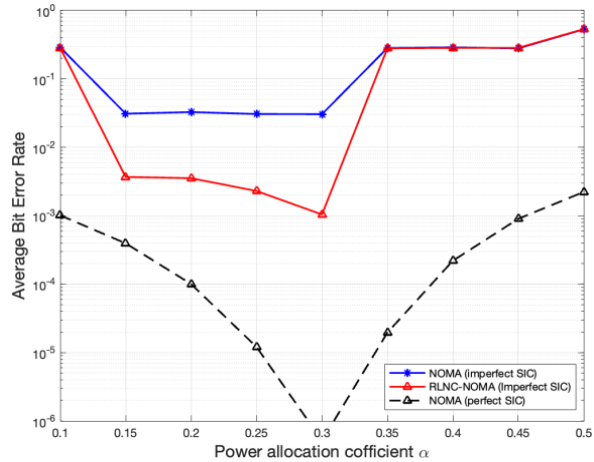

*Figure 4. The average BER of NOMA w/o RLNC versus power coefficient α.*

## 4. CONCLUSIONS

In this work, RLNC is applied to enhance the performance of NOMA in OWC systems. We defined a multicast system model that contains a single optical source with multiple users that differ in their channel gains. The users are divided into two multicast groups where each group contains users that slightly differ in their channel gains. After that, the average BER is evaluated in perfect and imperfect NOMA scenarios and the ergodic sum rates of RLNC for both NOMA and OMA schemes are analysed. The results show that the proposed RLNC-NOMA scheme provides improved performance compared to traditional NOMA and OMA schemes. Furthermore, its sensitivity to the power allocation coefficient is very high in all scenarios.

## ACKNOWLEDGEMENTS

The authors would like to acknowledge funding from the Engineering and Physical Sciences Research Council (EPSRC) INTERNET (EP/H040536/1), STAR (EP/K016873/1) and TOWS (EP/S016570/1) projects. For the purpose of open access, the authors have applied a Creative Commons Attribution (CC BY) licence to any Author Accepted Manuscript version arising. All data are provided in full in the results section of this paper.